\documentclass[12pt]{article}
\usepackage{graphicx}
\usepackage{subfig}
\usepackage{booktabs}
\usepackage{multirow}
\usepackage{color}
\usepackage{wrapfig}
\usepackage{adjustbox}
\usepackage{placeins}
\usepackage[%
font={small},
labelfont=bf,
format=hang,    
format=plain,
margin=0pt,
width=\textwidth,
]{caption}
\usepackage{tablefootnote}
\usepackage{threeparttable}

\begin{document}

\definecolor{ForestGreen}{RGB}{63,142,38}
\definecolor{UnibasMint}{RGB}{30,165,165}

\newcommand{\aliA}[1]{{\color{black}#1}}

\newcommand{\flo}[1]{{\color{black}#1}}
\newcommand{\discuss}[1]{{\color{black}#1}}
\newcommand{\ali}[1]{{\color{black}#1}}

\newcommand{\simdag}{\mbox{SG-SD}}
\newcommand{\msg}{\mbox{SG-MSG}}
\newcommand{\smpi}{\mbox{SG-SMPI}}
\newcommand{\simgrid}{\mbox{SG}}

%FLO: Enclose in \cut{text to be left out}{} the text to be left out for the longer version / journal extension
\newcommand{\cut}[1]{}

%\title{Examining and Enhancing the Performance of Dynamically Scheduled Scientific Applications on Heterogeneous High Performance Computing Systems in the Presence of Perturbations\\
%(How to select the most appropriate DLS? SiL)}
\title{SiL: An Approach for Adjusting Applications to Heterogeneous Systems\\ Under Perturbations}
%\titlerunning{SiL: An Approach for Adjusting Applications to Heterogeneous Systems}
\author{Ali Mohammed and Florina M. Ciorba\\
	Department of Mathematics and Computer Science\\
	University of Basel, Switzerland\\
}	
%\author{Ali Mohammed \and Florina M. Ciorba}
%
%\authorrunning{A. Mohammed et al.}
%\institute{Department of Mathematics and Computer Science\\
%University of Basel, Switzerland\\
%\email{\{firstname.lastname\}@unibas.ch}}
%
\maketitle              % typeset the header of the contribution
\clearpage
\setcounter{tocdepth}{2}
\tableofcontents
\clearpage

% !TEX root =  2018_dls_perturbations.tex
\vspace{-0.5cm}
\begin{abstract}
\label{sec:abstract}

%The abstract should briefly summarize the contents of the paper in 15--250 words.

%elements
%1. A problem statement of the research under consideration.
Scientific applications consist of large and computationally-intensive loops. 
\cut{, such as N-body, Monte Carlo, and computational fluid dynamics }%
%These loops contain computationally-intensive operations, resulting in heavy loop bodies. 
%Loop scheduling techniques are used to parallelize such applications. 
%Dynamic loop scheduling (DLS) techniques are used to achieve \ali{a balanced load} execution of such applications.
Dynamic loop scheduling (DLS) techniques are used to load balance the execution of such applications. 
Load imbalance can be caused by variations in loop iteration execution times due to problem, algorithmic, or systemic characteristics (also, perturbations).
\cut{Therefore, DLS achieve a balanced load execution of scientific applications on high-performance computing systems~(HPC).}%
\cut{The variations in systemic characteristics are referred to as perturbations, and can be caused by other applications or processes that share the same resources, or a temporary system fault or malfunction and include, decreased delivered computational speed, reduced available network bandwidth, or larger network latencies.}%Motivation 
The following question motivates this work:~\emph{``Given an application, a high-performance computing~(HPC) system, and both their characteristics and interplay, which DLS technique will achieve improved performance under unpredictable perturbations?''}%in different execution scenarios 
%%Existing solutions
%Existing studies focus on variations in the delivered computational speed only as source of perturbations in the system.
\cut{Scheduling solutions based on optimization techniques, e.g., evolutionary algorithms, can not adapt to perturbations during execution.}
\cut{The alternative of using machine learning for DLS selection requires training and learning either prior to execution or during previous time-steps in time-stepping applications.}%
Existing work only considers perturbations caused by variations in the \aliA{HPC system delivered computational speeds}.
However, perturbations in available network bandwidth or latency are inevitable on production HPC systems. 
\emph{Simulator in the loop} (SiL) is introduced, herein, as \aliA{a new control-theoretic inspired approach} to dynamically select DLS techniques that improve the performance of applications on \aliA{heterogeneous HPC systems} under perturbations.
The present work examines the performance of six applications on a heterogeneous system under \emph{all above system perturbations}.\cut{in the delivered computational speed, available network bandwidth and latency.}
The SiL \emph{proof of concept} is evaluated using simulation.\cut{the SimGrid simulation toolkit.}
The performance results confirm the initial hypothesis that \emph{no single DLS technique can deliver best performance in all scenarios}, while the SiL-based DLS selection delivered improved application performance in most experiments.
\vspace{-0.25cm}
%\keywords{
\end{abstract}
\clearpage
%

% !TEX root =  2018_dls_perturbations.tex
\section{Introduction}
\label{sec:intro}

Scientific applications are often characterized by large and computationally-intensive parallel loops. 
%The loop iterations in such applications may have an irregular number of computations.
The \ali{performance} of these applications on high-performance computing (HPC) systems may degrade due to load imbalance caused by problem, algorithmic, or systemic characteristics. 
%Variations in loop iteration execution times can be caused by 
Application (problem or algorithmic) characteristics include \ali{the irregularity of the} number of computations per loop iterations due to conditional statements, \ali{where} systemic characteristics include variations in delivered computational speed of processing elements~(PEs), available network bandwidth or latency.
Such variations are referred to as perturbations, and can also be caused by other applications or processes that share the same resources, or a temporary system fault or malfunction.
%\flo{Just to be clear: the way you have now defined perturbations, they strictly refer to systemic variations. The problem or algorithmic variations are *not* in. If we want to study a combination of systemic variations (perturbations) and problem+algorithmic variations, we will not be able to call them *all* perturbations. Right?}
%
Dynamic loop scheduling (DLS) is a widely-used approach for improving the execution of parallel applications using self-scheduling, that is \emph{dynamic} assignment of the loop iterations to free and requesting processing elements.
A wide range of DLS techniques exists, and can be divided into \emph{nonadaptive} and \emph{adaptive} techniques. 
The nonadaptive DLS techniques account for the variability in loop iterations execution times due to application characteristics. 
They do not account for irregular system characteristics that are known only during execution.  
The nonadaptive DLS techniques include \emph{self-scheduling}~(SS), \emph{fixed size chunking}~(FSC)~\cite{FSC}, \emph{guided \mbox{self-scheduling}}~(GSS)~\cite{GSS}, \emph{factoring}~(FAC)~\cite{FAC}, and \emph{weighted factoring}~(WF)~\cite{WF}.  
The adaptive DLS techniques account for irregular system characteristics by adapting the amount of assigned work per PE request (chunk size) according to the application performance measured during execution.
Adaptive DLS techniques include \emph{adaptive weighted factoring}~(AWF)~\cite{AWF}, its variants \emph{batch}~(AWF-B), \emph{chunk}~(AWF-C), \emph{batch-like}~(AWF-D), \emph{chunk-like}~(AWF-E)~\cite{AWFBC}, and \emph{adaptive factoring}~(AF)~\cite{AF}. 

An \emph{a priori} selection of the most appropriate DLS technique for a given application and system is challenging, given the various sources of load imbalance and the different load balancing properties of the DLS techniques. 
This observation raises the following question and motivates the present work: \emph{``Given an application, an HPC system, and both their characteristics and interplay, which DLS technique will achieve improved performance under unpredictable perturbations?''}
Earlier work studied the flexibility of DLS~(robustness to reduced delivered computational speed)~\cite{sukhija:2013:b} and the selection of robust DLS using machine learning~\cite{sukhija:2014:a} \aliA{with} the SimGrid~(\simgrid{})~\cite{SimGrid} simulation toolkit.
The selection of DLS techniques for synthetic time-stepping scientific applications using reinforcement learning~\cite{Boulmier:2017a} was also studied using \simgrid{}.\cut{ where smart agents learn from application performance in previous time-steps to select a DLS that would improve the performance.} 
The aforementioned existing work focuses on one source of perturbations (variation in delivered computing speed) in time-stepping applications to learn from previous steps. 
That approach may not be applicable to applications without time-steps, nor would it be feasible in a highly variable execution environment.
Scheduling solutions using static optimizations, e.g., using evolutionary \ali{and genetic} algorithms, can not dynamically adapt to the perturbations encountered during execution.
Modern HPC systems are often heterogeneous production systems typically shared by many users.
Therefore, perturbations in the available network bandwidth and latency in such systems are unavoidable. 

In the present work, in an effort to select the most appropriate DLS for a given application and system, the performance of a scientific application~(PSIA~\cite{Eleliemy:2017b}) and five synthetic applications using nonadaptive and adaptive DLS techniques is studied on a heterogeneous HPC system, in the presence of perturbations in computing speed, network bandwidth, and network latency. 
%The selected applications represent a wide range of scientific applications. 
The amount of operations in each loop iteration \ali{of the five synthetic applications} is assumed to follow five different probability distributions, namely: constant, uniform, normal, exponential, and gamma probability distributions.
\cut{This assumption captures the characteristics of a wide range of applications.}
The present work makes the following contributions:
(1)~Proposes a novel \emph{simulator in the loop} (SiL) approach for dynamically selecting a DLS technique during execution, based on the application characteristics and the present (monitored or predicted)\cut{ (or predicted for the near future)} state of the computing system; 
(2)~Provides insights on the resilience of the DLS techniques to perturbations; and % from the performance analysis; 
(3)~Confirms the initial hypothesis that no single DLS ensures the best performance in all execution scenarios considered;
The SiL performance is evaluated for the selected applications in simulation using \simgrid{}. 
%\flo{The implementation of DLS techniques in \simgrid{} has been experimentally verified in a previous study~\cite{Mohammed:2018a}. - consider moving it only to rel. work. }

This work is structured as follows. 
Section~\ref{sec:background} contains a brief review of the selected DLS techniques, the \simgrid{} simulation toolkit, as well as of the work related to the performance of scheduling scientific applications with DLS in the presence of perturbations. 
The proposed SiL approach for selecting a DLS technique in the presence of perturbations \ali{is} discussed in Section~\ref{sec:sil}. 
The experimental design and setup, and the performance of the proposed approach are described and discussed in Section~\ref{sec:evalaution}. 
\cut{(applications, DLS, HPC system, and perturbations characteristics),}
%The performance of the proposed approach on a heterogeneous HPC system under perturbations is presented and examined in Section~\ref{sec:evalaution}. 
The work concludes and outlines potential future work in Section~\ref{sec:conc}.

%\begin{itemize}
%	\item A general overview of the field
%	\item The problem statement
%	\item Existing solutions and their criticism should be limited normally to only those directly relevant to the contribution
%	\begin{itemize}
%		\item intro into SimGrid -- using SimGrid to study DLS resilience and flexibility~\cite{sukhija:2013:b}  and \cite{sukhija:2015} 
%		\item simulate app into SimGrid ~\cite{Ali_SC:17} -- how realistic is simulation~\cite{Mohammed:2018a}
%		\item Autonomic DLS selection~\cite{Boulmier:2017a},its contributions and limitations
%	\end{itemize}
%	\item Proposed method
%	\item Contributions
%	\item Conditions, context, assumptions, and limitations of the research
%	\item The structure and content of the rest of the document
%\end{itemize}

%

% !TEX root =  2018_dls_perturbations.tex
\section{Background and Related Work}
\label{sec:background}
\cut{In this section, a background on loop scheduling techniques and the \simgrid{} simulation toolkit is provided. Certain relevant work on robust loop scheduling is presented and discussed.}
 
\textbf{Loop scheduling.}
%Loop scheduling is the assignment of loop iterations in space (to PEs) and in time. 
The aim of loop scheduling is to achieve a balanced load execution among the parallel PEs with minimum scheduling overhead.
Loop scheduling can be divided into \emph{static} and \emph{dynamic}. 
In static loop scheduling, the loop iterations are divided and assigned to PEs before execution; both division and assignment remain fixed during execution. 
%Examples of static loop scheduling include block, cyclic, and block-cyclic. 
This work considers static (block) scheduling, denoted STATIC, each PE being assigned a chunk size equal to the number of iterations $N$ divided by the number of PEs $P$.
STATIC incurs \emph{minimum} scheduling overhead, compared to dynamic loop scheduling, and may lead to load imbalance for non-uniformly distributed tasks and/or on perturbed systems.

In \emph{dynamic loop scheduling}~(DLS), free and requesting PEs are assigned, via self-scheduling, loop iterations during execution. 
The DLS techniques can be categorized into \emph{nonadaptive} and \emph{adaptive} techniques. 
The nonadaptive DLS techniques considered in this work are: SS~\cite{SS}, FSC~\cite{FSC}, GSS~\cite{GSS}, FAC~\cite{FAC}, and WF~\cite{WF}.
While STATIC represents one scheduling extreme, SS represents the other scheduling extreme.
In SS, \ali{the} size of each chunk is one loop iteration.
This yields a high load balance with potentially very large scheduling overhead.
%FSC tries to minimize the scheduling overhead and maximize load balance. %between the large overhead of SS and the large chunk size of STATIC and its poor load balancing. 
\ali{FSC} assigns loop iterations in chunks of fixed sizes, where the chunk size depends on the standard deviation of loop iteration execution times~$\sigma$ as an indication of its variation and the incurred scheduling overhead~$h$.
%FSC requires this information ($h$ and $\sigma$) to be known before the execution to calculate the chunk size.
GSS assigns loop iterations in chunks of decreasing sizes, where the size of a chunk is equal to the number of remaining unscheduled loop iterations $R$ divided by the number of PEs $N$.
%FAC assigns chunks of iterations in batches to minimize the number of chunk calculation operations, where the chunk size is equal to the batch size divided by $N$.  
FAC employs a probabilistic modeling of loop characteristics (standard deviation of iterations execution time $\sigma$ and their mean $\mu$) to calculate batch sizes that maximize the probability of achieving a load balanced execution.
\flo{A PE's chunk size is equal to the batch size divided by $N$.}
When this information ($\sigma$ and $\mu$) is unavailable, FAC is practically implemented to assign half of the remaining loop iterations $R$ in a batch. 
%WF is an extension of FAC where it takes into its account the heterogeneity of the computing system. 
WF divides a batch into unequally-sized chunks, proportional to the relative PE speeds (weights).
The PEs weights need to be determined prior to the execution and do not change afterward.
This work considers the \emph{practical implementations} of FAC and WF.
All nonadaptive DLS techniques account for variations in iteration execution times due to application characteristics. 
%However, they do not adapt the chunk calculation to accommodate the variability in the execution due to unforeseen system characteristics during execution.     

The adaptive DLS techniques measure the performance of the application during execution and adapt the chunk calculation accordingly. % to achieve balanced execution.
Adaptive DLS techniques include AWF~\cite{AWF}, its variants~\cite{AWFBC}: AWF-B, AWF-C, AWF-D, AWF-E, and AF~\cite{AF}.
AWF is designed for time-stepping applications.
It improves WF by changing the relative weights of PEs during execution by measuring their performance in each time step and updating their weights accordingly.
AWF-B relieves the time stepping requirement in AWF, and measures the performance after each batch to update the PE weights.
AWF-C is similar to AWF-B, where weight updates are performed upon the completion of each chunk, instead of a batch.
AWF-D is similar to AWF-B, and considers the total chunk time (equal to the chunk iteration execution times plus the associated overhead of a PE to acquire the chunk) and all the bookkeeping operations to calculate and update the PE weights. 
AWF-B and AWF-C only consider the chunk iterations execution times.
AWF-E is similar to AWF-C by updating the PE weights on every chunk. 
Yet AWF-E is also similar to AWF-D by also considering the total chunk time also.
%AF is an evolution of FAC, where the constraint of known $\sigma$ and $\mu$ is relieved.
Unlike FAC, \ali{AF} dynamically estimates the values of $\sigma$ and $\mu$ during execution and updates them based on the measured performance of the PEs.
% 
%\textbf{Static and dynamic}\\
%STATIC and SS~\cite{SS} the two extremes.\\
%dynamic is nonadaptive and adaptive.\\
%\textbf{nonadaptive}\\
%SS, FSC~\cite{FSC}, GSS~\cite{GSS}, FAC~\cite{FAC}, WF~\cite{WF}\\
%\textbf{adaptive}
%AWF(-B,-C,-D,-E)~\cite{AWFBC} and AF~\cite{AF}
 
\emph{Loop scheduling in simulation.}
SimGrid~\cite{SimGrid}~(\simgrid{}) is a versatile event-based simulation toolkit designed for the study of the behavior of large-scale distributed systems. % such as Grids, Clouds, HPC or P2P systems. 
It provides ready to use application programming interfaces~(API) to represent applications and computing systems through different interfaces: MSG~(\msg{}), SimDag~(\simdag{}), and SMPI~(\smpi{}).
%
%The \smpi{} interface is intended to study the performance of MPI applications, and it accepts native MPI application codes for simulation. 
%The \msg{} interface is designed to represent parallel applications that are represented as parallel communicating processes. 
%The \simdag{} interface is for simulating applications represented as a directed acyclic graph (DAG) of tasks. 
%
\simgrid{} uses a simple, fast CPU computation model and verified network models~\cite{velho2009accuracy} which render it well suited for the study of computationally-intensive distributed scientific applications.

Various studies have used \simgrid{} to study the performance of applications with DLS techniques in different scenarios~\cite{Boulmier:2017a,sukhija:2013:b,sukhija:2014:a}.
\ali{To attain high trustworthiness in the performance results obtained with \simgrid{},} 
the implementation of the nonadaptive DLS techniques in \simdag{} has been verified~\cite{HPCS} by reproducing the results presented in the work that introduced factoring~\cite{FAC}.
\ali{Also, the accuracy of simulative performance experiments against native experiments has recently been quantified}~\cite{Mohammed:2018a}.
%, the percent error between native and simulative performance results was found to be between $0.95\%$ and $2.99\%$ for the \simdag{} interface simulation the performance of PSIA~\cite{Eleliemy:2017b} with different DLS techniques on an HPC platform.
This work employs the \simdag{} interface to study the performance of scientific applications on a heterogeneous platform under perturbations. 

%Introduction~\cite{SimGrid}, interfaces, why SimGrid ...suitable for the study of computationally intensive distributed scientific applications, simple CPU model, verified network models~\cite{velho2009accuracy}. Why SimDag .. represent tasks with and without dependencies.
%
%\textbf{SimGrid and DLS} verified DLS implementation in MSG~\cite{hoffeins2017examining} and in SimDag~\cite{Ali_HPCC:17}. Used in many studies, such as~\cite{sukhija:2013:b} for flexibility,~\cite{sukhija:2014:a} and~\cite{Boulmier:2017a}. 
\vspace{0.5cm}
\textbf{Related work.}
%Many studies have examined the performance of applications with DLS in the presence of perturbations. 
\ali{Robustness denotes} the maintenance of certain desired system characteristics despite fluctuations in the behavior of its components or its environment~\cite{ali2004measuring}, whereas, flexibility~\cite{sukhija:2013:b} \ali{denotes the} robustness of DLS to variations in the delivered computational speeds.
The performance of scientific applications under perturbations in the delivered computational speed is studied with nonadaptive DLS techniques~\cite{DBLP:conf/parco/Garcia-Gonzalez17,yang2003rumr}. 
The robust scheduling of tasks with uncertain communication time was \ali{also} considered using a multi-objective evolutionary algorithm~\cite{canon2010evaluation} and to evaluate the flexibility of DLS~\cite{sukhija:2013:b}.
The selection of the best performing DLS during execution was studied for OpenMP multi-threaded applications~\cite{zhang2005runtime}, and for time-stepping applications using reinforced learning~\cite{Boulmier:2017a}.
Also, machine learning was used to create a portfolio of DLS robustness to variations in the delivered computational speed on a homogeneous system~\cite{sukhija:2014:a}.

Scheduling solutions based on optimization techniques, e.g., genetic and evolutionary algorithms, can not adapt to perturbations during execution.
%The alternative of using machine learning for DLS selection requires training and learning either prior to execution or during previous time-steps in time-stepping applications. {%to conclusion cite antony and nitin SiL cost to machine learing}
\ali{None of the aforementioned efforts} considered perturbations in network bandwidth and latency. % which is inevitable on shared heterogeneous HPC systems.  
\ali{This work complements the previous efforts} by studying the performance of scientific applications using nonadaptive and adaptive DLS techniques under different perturbations (variations in delivered computational speed, network bandwidth, network latency) \ali{on} a heterogeneous computing system.
A new approach, namely \emph{simulator in the loop} (SiL) is introduced, to dynamically select DLS techniques that improve the performance of applications on heterogeneous system under multiple sources of perturbations.
\section{Simulator in the Loop (SiL)}
\label{sec:sil}
%Even though adaptive DLS techniques are robust and result in a good performance in the presence of perturbation, other DLS techniques may outperform them in a certain execution scenario. 
%In this section, the simulator in the loop (SiL) approach is presented. 
%\subsection{Proposed Approach}
%\label{subsec:approach}

\begin{figure}[]
%\vspace{-1cm}
	\begin{minipage}{0.4\textwidth}
			\label{subfig:control}%
		\centering
		%	\begin{adjustbox}{minipage=\linewidth,frame}
			\includegraphics[clip, trim=0cm 0cm 0cm 0cm,scale=0.4]{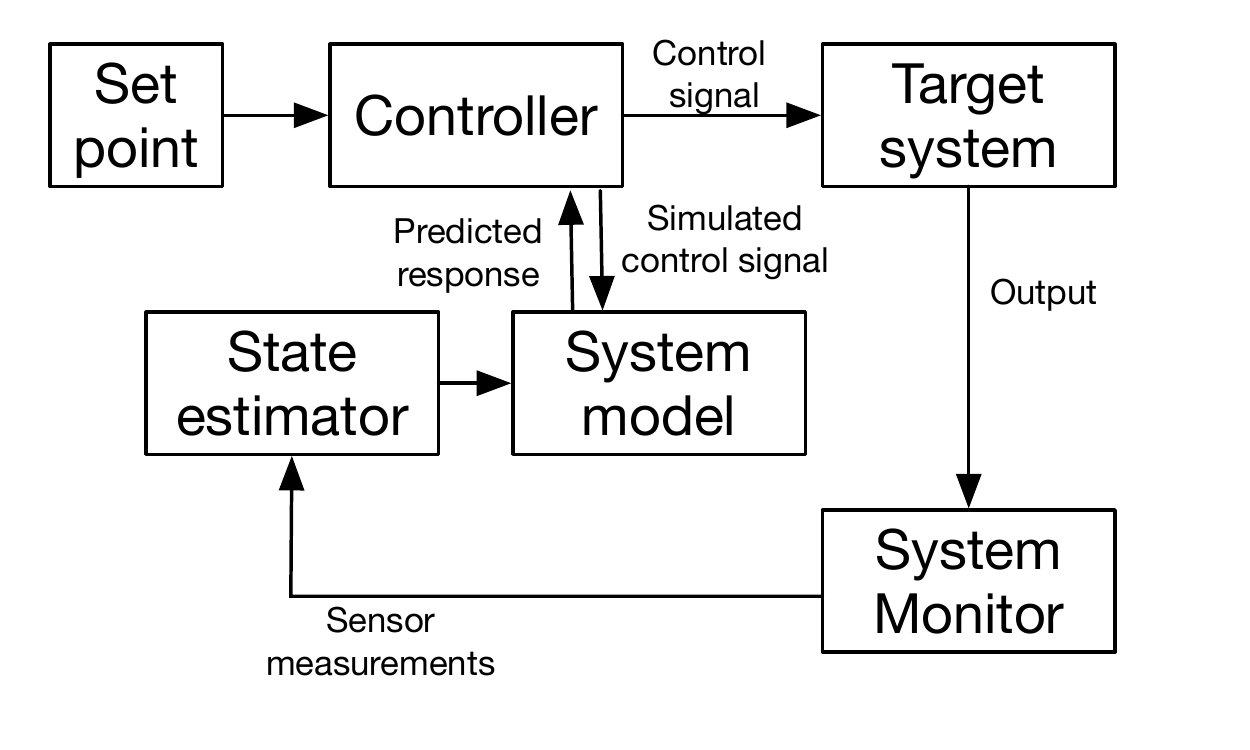}\\
			\small{(a) A generic control system.}

	\end{minipage}
      \begin{minipage}{0.6\textwidth}
      	\label{subfig:sil}%
      		\includegraphics[clip, trim=0cm 0cm 0cm 0cm,scale=0.44]{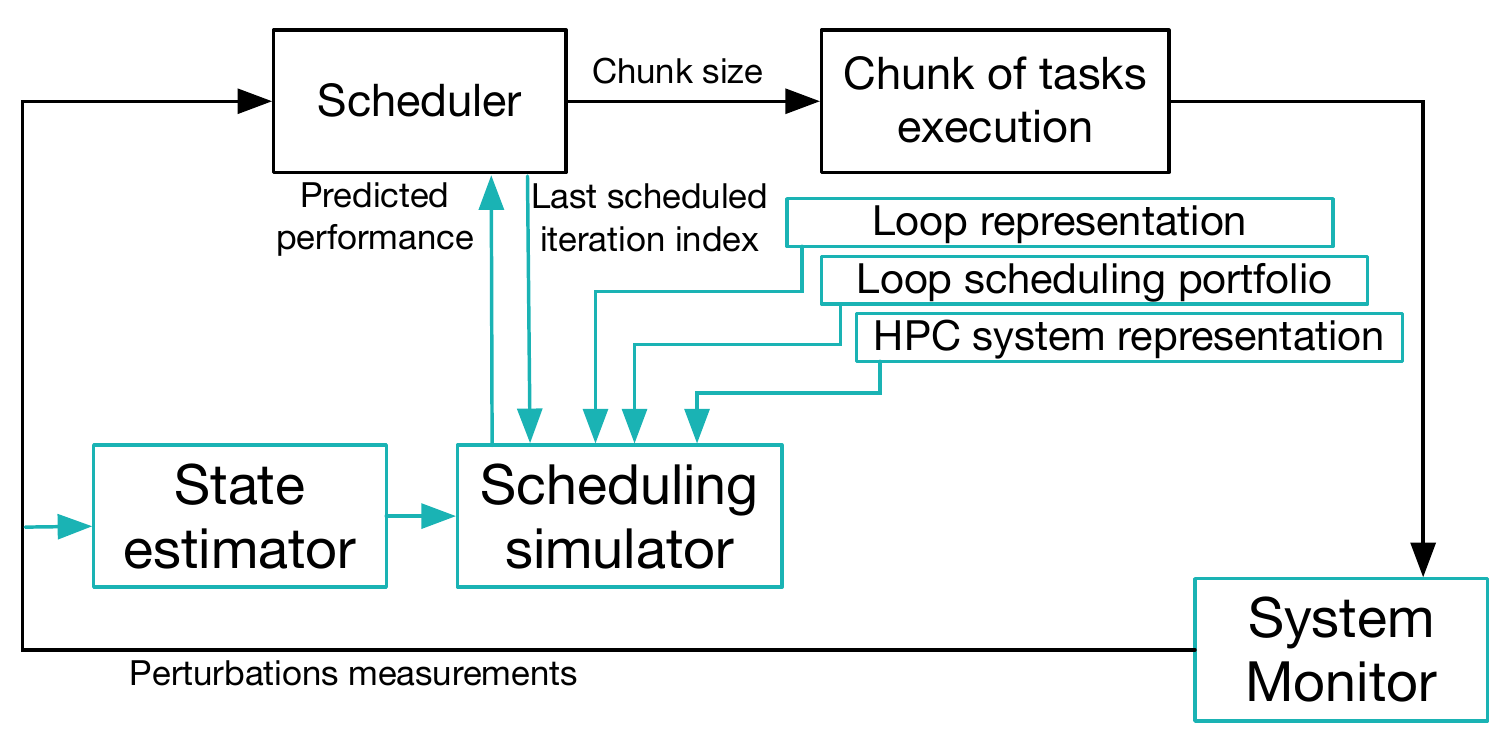}\\
      		\small{(b) Proposed SiL approach for loop scheduling.}
      		
      \end{minipage}%\hspace{0cm}
		 %\vspace{-0.25cm}
		\\	
		\caption{The proposed \emph{simulator in the loop} (SiL) approach for loop scheduling (b) is analogous to a typical control system (a). The components highlighted in mint color in (b) represent the SiL additions to a typical loop scheduling system.} %\flo{How do you define state in both (a) and (b) subfigures? They are not defined and they need to be.}}
		\label{fig:approach}
%\vspace{-.75cm}
\end{figure}

The SiL is inspired by control theory, where a controller (scheduler) is used to achieve and maintain a desired state (load balance) of the system (parallel loop execution), as illustrated in \figurename{~\ref{fig:approach}}.
% by an analogy to the structure of a control system.
The SiL concept is motivated by the well-known control strategy model predictive control~(MPC)~\cite{rawlings2000tutorial}. %, where different control signals are tested on a model of the controlled system before it is applied to the real control system. 
The MPC~controller predicts the performance of the system with different control signals to optimize system performance.
As shown in \figurename{~1(b)}, a call to the SiL simulator is inserted inside \ali{a} typical scheduling loop. 
SiL leverages state-of-the-art simulation toolkits to estimate the performance of an application in a given execution scenario.
%, to predict the performance of the application with various DLS techniques and recommends the DLS technique that would result in the best performance, to the application. 
%The simulator takes in the parallel loop characteristics and the computing system representation.
The system monitor and estimator components read the system state during the execution and update the computing system representation accordingly.
The above steps may \aliA{be repeated} several times during the execution of the loop, and this frequency can be aligned with the perturbations frequency or intensity.

%The simulator simulates the application execution, starting from iteration $i$ (read from the application) and simulates the application execution using the updated system representation with a portfolio of DLS techniques that are supported by the simulator and the application.
%The DLS technique that results in the best performance is then selected and used by the application. 
%
%
%\begin{figure}[]
%	\centering
%	\includegraphics[clip, trim=0cm 1.5cm 0cm 0cm,scale=0.6]{figures/approach.pdf}%
%	\caption{The proposed simulator in the loop (SiL). The simulator is called every period ($T$), where ($t$) is the current time. The HPC representation is updated dynamically by system monitor and predictor module to reflect the latest perturbations.}
%	\label{fig:approach}
%\end{figure}
The advantage of SiL is that it leverages the use of already developed state-of-the-art simulators to predict the performance dynamically during execution. 
The accuracy of the simulator and its prediction is strongly influenced by the representation of both applications and the systems in simulation as well as by the available subsystems models in the simulator~\cite{Mohammed:2018a}. 
For instance, the percent error between native and simulative executions for a given application (PSIA~\cite{Eleliemy:2017b}) using the \simdag{} interface was found to be between $0.95\%$ and $2.99\%$~\cite{Mohammed:2018a}.
It is expected that the accuracy and the speed of the simulators employed by SiL will improve as \aliA{they are continuously being developed and refined}. 
The cost of frequent calls to SiL can be amortized by launching parallel SiL instances to concurrently derive predictions for various DLS.
Alternatively, this cost can be entirely mitigated by asynchronously calling SiL, concurrently to the application execution.
Upon completion, SiL returns the recommended \emph{best suited DLS technique} to the calling application, which can then \aliA{directly} use the recommended DLS to improve the application performance. 

The system monitor and estimator components can be implemented with a number of system monitoring tools~\cite{Ciorba:2016}, such as \texttt{collectl}. 
\aliA{Such} tools can periodically be instantiated to measure PE and network loads and to update the system representation in the simulator.
The measured chunk execution times can also be used to estimate the current PE computational speeds.
\cut{An overview of system monitoring and tools can be found in the literature~\cite{Ciorba:2016}.}
The PE loads can be estimated and predicted using autoregressive integrated moving average~\cite{mehrotra2015power}.\cut{where a control theory inspired method is introduced for power aware scheduling of tasks on heterogeneous systems.}

%
% !TEX root =  2018_dls_perturbations.tex
\section{Evaluation and Analysis}
\label{sec:evalaution}

\cut{A summary of the details of the experiments considered in this work is presented in Table~\ref{tbl:ex}. 
The performance results of the execution of the applications with different loop scheduling techniques under different execution scenarios are presented and discussed.}{}

\textbf{Experimental Design and Setup.}
\label{subsec:exp}
The factorial design of experiments is presented in the following (cf. Table~\ref{tbl:ex}), together with the applications performance and a discussion thereof.
\cut{The applications considered in this work and their representation in the simulation are presented.
The considered DLS techniques and how they are implemented are explained. 
Details of the heterogeneous HPC cluster used in this study and how it is simulated in the \simgrid{} are provided.
The considered perturbations scenarios are presented.}
\begin{table}[h]
%	\vspace{-0.5cm}
	\centering
	\caption{Design of factorial experiments}
	\label{tbl:ex}
	\resizebox{\textwidth}{!}{% 
		\begin{threeparttable}
		\begin{tabular}{lll}
			\toprule
			\textbf{Factors}                                           & \textbf{Values}                                                                                                      & \textbf{Properties}                                                                                                                                        \\ \midrule
			\multirow{2}{*}{  \begin{tabular}[c]{@{}l@{}} \\   \\  \\ \textbf{Applications}  \end{tabular} }                 &  Problem size &$N =$ 400,000 iterations\\ \cline{2-3}  & \begin{tabular}[c]{@{}l@{}}  PSIA \\Constant\\Uniform\\Normal\\Exponential\\Gamma \end{tabular} &    \begin{tabular}[c]{@{}l@{}}  $[5.9 \cdot 10^7$, $6.6 \cdot 10^7]$ FLOP per iteration\\$2.3 \cdot 10^8$ FLOP per iteration\\ $[10^3$, $7 \cdot 10^8]$ FLOP per iteration\\ $\mu = 9.5 \cdot 10^8$ FLOP, $\sigma= 7\cdot 10^7$ FLOP, $[6 \cdot 10^8, 1.3 \cdot 10^9 ]$ FLOP per iteration\\ $\lambda = 1/3 \cdot 10^8$ FLOP, $[948$, $4.5 \cdot 10^9]$ FLOP per iteration\\ $k = 2$, $\theta = 10^8$ FLOP, $[4.1 \cdot 10^6$, $2.7 \cdot 10^9]$ FLOP per iteration \end{tabular}      \\ \hline
			\textbf{Loop scheduling}                & \begin{tabular}[c]{@{}l@{}}STATIC\\ SS, FSC, GSS, FAC, WF\\ AWF-B, -C, -D, -E, AF\end{tabular}                     & \begin{tabular}[c]{@{}l@{}}Static\\ Nonadaptive dynamic\\ Adaptive dynamic\end{tabular}                                                                    \\ \hline
			\textbf{Computing system}               & \begin{tabular}[c]{@{}l@{}}miniHPC \\ (heterogeneous HPC cluster)\end{tabular}                                     & \begin{tabular}[c]{@{}l@{}}22 Intel Broadwell nodes ($22 \cdot 20$ cores), relative core weight $= 1.398$\\ 4 Intel Xeon Phi KNL nodes ($4 \cdot 64$ cores), relative core weight $= 0.316$\\ $P= 224$ heterogeneous (112 Broadwell + 112 KNL) cores\\ $P= 696$ heterogeneous (440 Broadwell + 256 KNL) cores \end{tabular} \\ \hline
			\multirow{5}{*}{  \begin{tabular}[c]{@{}l@{}} \\ \\ \\ \\ \\  \\  \\ \textbf{Perturbations}  \end{tabular}}  & Nominal conditions &  no perturbations (np)    \\ \cline{2-3} & PE availability                                                                                          & \begin{tabular}[c]{@{}l@{}} constant mild (pea-cm)\\  constant severe (pea-cs) \\ exponential mild (pea-em) \\ exponential severe (pea-es)\end{tabular}                                  \\ \cline{2-3}
			& Bandwidth                                                                                                       & \begin{tabular}[c]{@{}l@{}} constant mild (bw-cm)\\ constant severe (bw-cs)\\ exponential mild (bw-em)\\ exponential severe (bw-es)\end{tabular}                                                                          \\ \cline{2-3}
			& Latency                                                                                                         & \begin{tabular}[c]{@{}l@{}} constant mild (lat-cm)\\ constant severe (lat-cs)\\ exponential mild (lat-em)\\ exponential severe (lat-es)\end{tabular}        \\ \cline{2-3}
			& All                                                                                                             & \begin{tabular}[c]{@{}l@{}} constant mild (all-cm)\\ constant severe (all-cs)\\ exponential mild (all-em)\\ exponential severe (all-es)\end{tabular}                \\ \hline
				\multirow{2}{*}{  \begin{tabular}[c]{@{}l@{}} \textbf{Experimentation}  \end{tabular}} & Native\tnote{a} &  PSIA on $224$ cores under no perturbations (online\tnote{b} ) \\ \cline{2-3}
				& Simulative  &   \begin{tabular}[c]{@{}l@{}} All applications on $224$ cores under all perturbations (online\tnote{b} ) \\ All applications on $696$ cores under all perturbations  \end{tabular} \\ \bottomrule
		\end{tabular}%
	\begin{tablenotes}
		\item[a] Direct experiments on real HPC systems.
		\item[b] Included in this arxiv.org submission, please download all data 
	\end{tablenotes}
\end{threeparttable}
	}
%	\vspace{-0.65cm}
\end{table}

\textit{Applications.}
\label{paragraph:app}
This work considers a real-world application and five synthetic applications.
The parallel spin-image algorithm~\cite{Eleliemy:2017b}~(PSIA), is an application from computer vision. 
The PSIA is algorithmically load imbalanced and the computational effort of a loop iteration depends on the input data. 
The performance of PSIA has been studied in prior work~\cite{Eleliemy:2017b} and enhanced for a heterogeneous cluster by using nonadaptive DLS techniques.
The total number of PSIA loop iterations is 400,000.
To represent the PSIA in simulation, the number of floating point operations (FLOP) of each loop iteration is counted using PAPI~\cite{papi} counters.
In \simdag{}, each loop iteration is represented as a task~\cite{Mohammed:2018a,Ali_SC:17}. 
Each of the five synthetic applications contains 400,000~parallel loop iterations, similar to the PSIA.
The FLOP count in each loop iteration is assumed to follow five different probability distributions, namely: constant, uniform, normal, exponential, and gamma probability distributions. 
%This assumption captures the characteristics of a wide range of applications.
The probability distribution parameters used to generate these FLOP counts are given in Table~\ref{tbl:ex}.

\textit{Loop scheduling.}
\label{paragraph:dls}
Eleven loop scheduling techniques are used to assess the performance of the above six applications under test. 
These techniques represent a wide range of loop scheduling approaches, namely, \emph{static} and \emph{dynamic}.
The dynamic loop scheduling (DLS) approach can further be distinguished into adaptive and nonadaptive.
The DLS techniques can be implemented using centralized or decentralized execution and control approach. 
The decentralized control approach was found to scale better by eliminating a centralized master, and hence, the master-level contention~\cite{HPCS}.
The DLS implemented using the decentralized control approach is considered in this work.

\textit{Computing system.}
\label{paragraph:comp}
%The computing system considered in this work is the miniHPC. 
\emph{miniHPC}\footnote{miniHPC is a fully controlled non-production HPC cluster at the Department of Mathematics and Computer Science at the University of Basel, Switzerland.} consists of 26~compute nodes: 22~nodes each with one dual socket Intel Xeon E5-2640~v4~(20~cores) configuration and 4 nodes each with one Intel Xeon Phi Knights Landing 7210~processor~(64~cores). 
The total number of heterogeneous cores is $22$~nodes~$\times~20$~cores per node $+~4$ nodes $\times~64$ cores per node $=~696$ cores.
All nodes are interconnected with Intel Omni-Path fabrics in a nonblocking two-level fat-tree topology.

\textit{Simulation.}
\label{paragraph:sim}
A computing system is represented in \simgrid{} via an XML file denoted as \texttt{platform file}.
\simgrid{} registers each processor core from their representation as a \texttt{host} in the \texttt{platform file}.
The computational speed of a processor core is estimated \ali{by measuring a loop execution time and dividing it by the total number of floating point operations included in the loop}~\cite{Mohammed:2018a}.
A Xeon core was found to be four times faster than a Xeon Phi core \ali{as indicated by the relative core weights~(cf. Table~\ref{tbl:ex})}.
The network bandwidth and latency represented in the \texttt{platform file} are calibrated with the \simgrid{} calibration procedure\footnote{http://simgrid.gforge.inria.fr/contrib/smpi-calibration-doc/}.%~\cite{simgrid_calib}.

\textit{Perturbations.}
\label{paragraph:per}
Three different categories of perturbations are considered in this work, namely \emph{delivered computational speed}, \emph{available network bandwidth}, and \emph{available network latency}. 
Two intensities are considered, mild and severe, for each category. 
Two scenarios are considered for each intensity, where the value of the delivered computational speed is either constant or exponentially distributed. 
%\flo{What is the second scenario?}
%
All perturbations (cf. Table~\ref{tbl:ex})  are considered to occur periodically, with a period of $100$~seconds where the perturbations affect the system only during $50\%$ of the perturbation period.
The network (bandwidth and latency) perturbations commence with the application execution, while the delivered computational speed perturbations begin 50~seconds after the start of the application.
The PE availability to compute changes to $75\%$ and $25\%$ for the mild and severe intensities, respectively.
The available network bandwidth and network latency change to $0.001\%$ and $0.00001\%$ for the mild and severe intensities, respectively.
Another perturbation scenario is created by combining all perturbations from the other individual categories.
%\aliA{The PE perturbations are applied to $50\%$ of the total number of cores, while the network perturbations act on all the links in the system.}
%The list of all considered perturbation scenarios is presented in Table~\ref{tbl:ex}.
All perturbations are enacted in \simgrid{} during simulation via the \texttt{availability}, \texttt{bandwidth}, \texttt{latency}, and \texttt{platform} files. \cut{ to represent perturbations in delivered computational speed, network available bandwidth, and network latency, respectively.}

\begin{figure}[]
	\begin{minipage}{1.2\textwidth}
		\centering
		%	\begin{adjustbox}{minipage=\linewidth,frame}
		\centering
		\vspace{-0.5cm}
		\subfloat[PSIA on 696 cores]{%
			\includegraphics[clip, trim=0cm 0.5cm 0cm 3.9cm, scale=0.36]{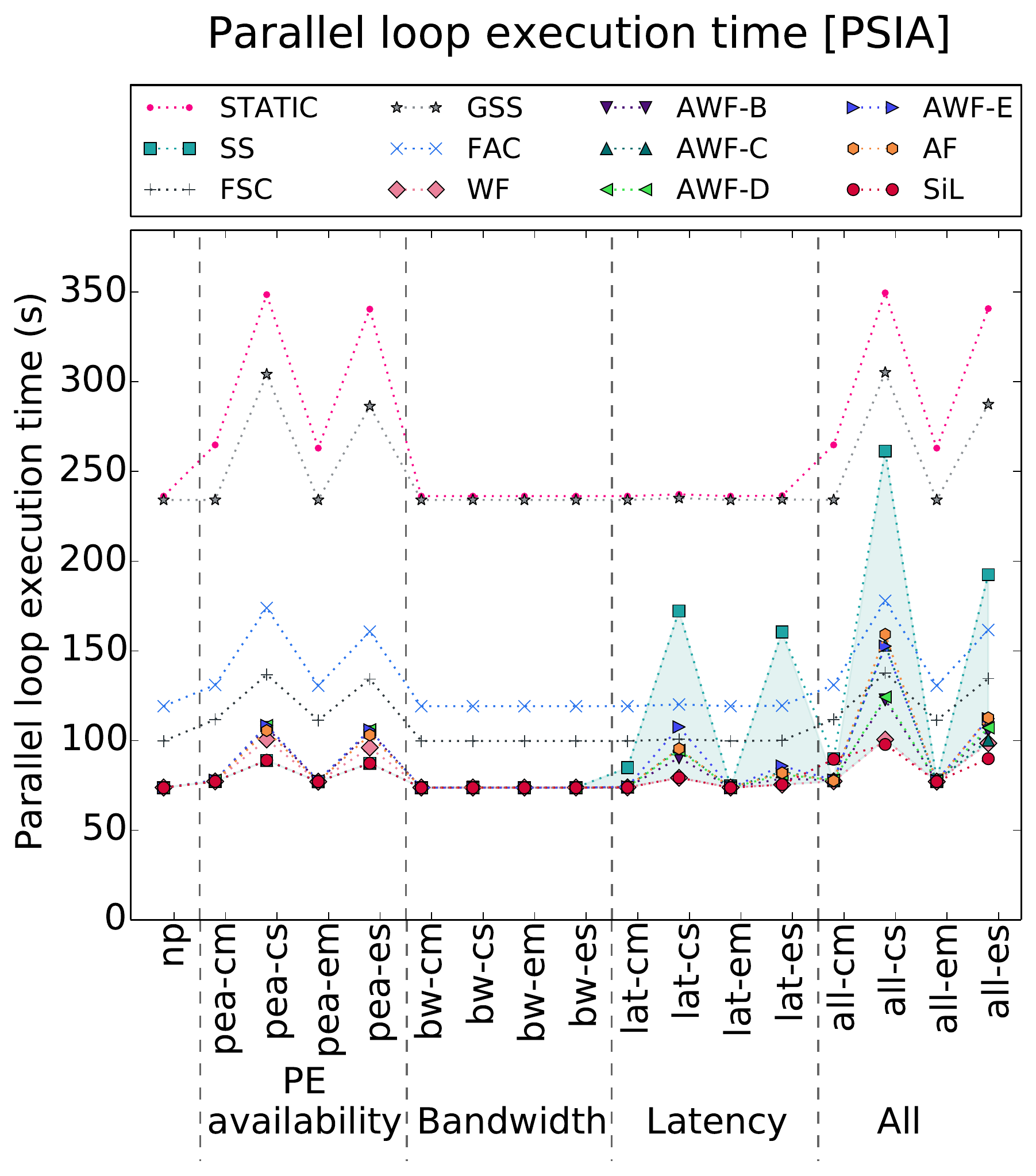}%
			\label{subfig:PSIA}%
		} \hspace{0cm} 
		\subfloat[Constant distribution on 696 cores]{%
			\includegraphics[clip, trim=0cm 0.5cm 0cm 3.9cm,scale=0.36]{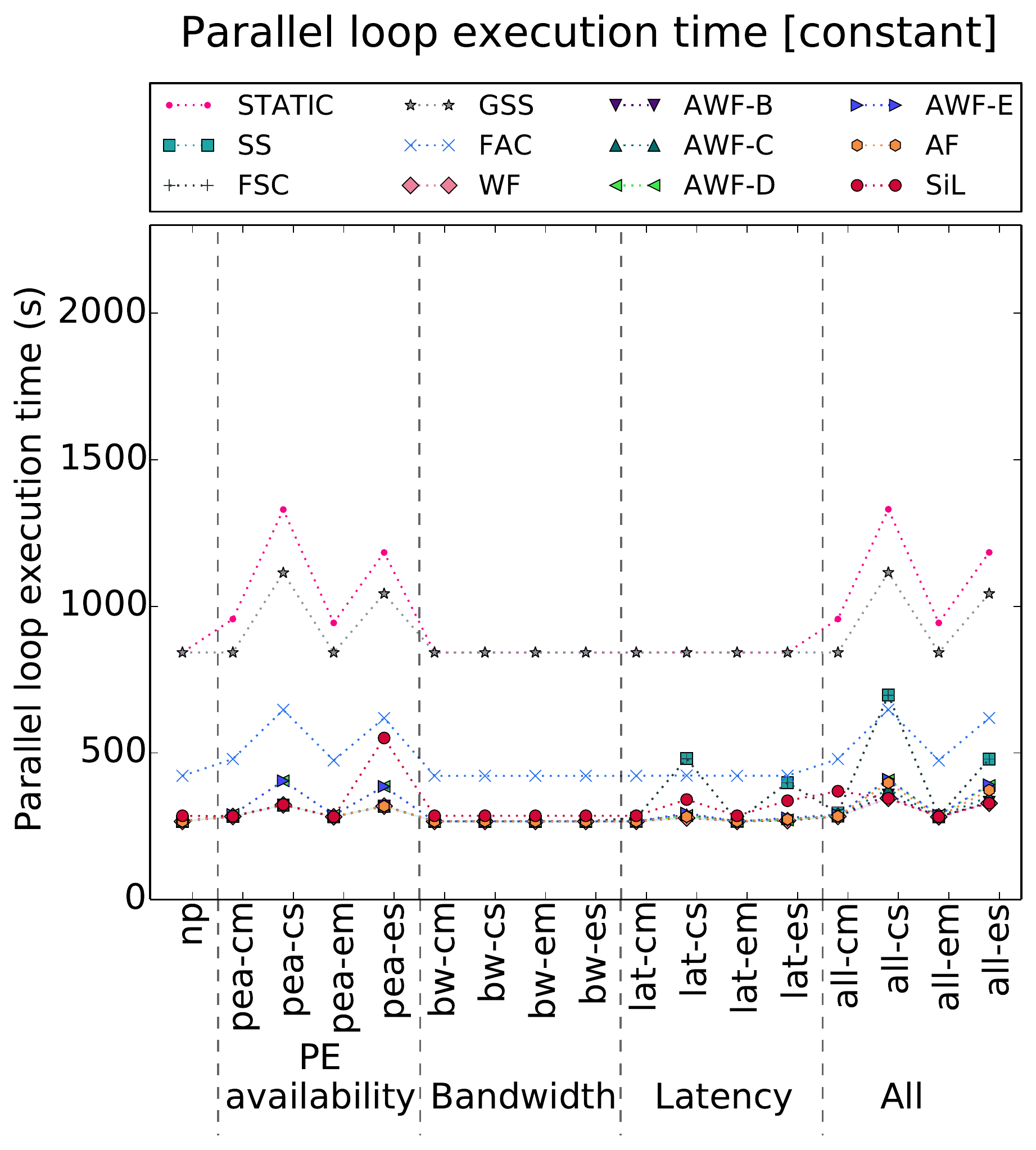}%%
			\label{subfig:constant}%
		} \vspace{-0.25cm}
		\\	
		\subfloat[Uniform distribution on 696 cores]{%
			\includegraphics[clip, trim=0cm 0.5cm 0cm 3.9cm,scale=0.36]{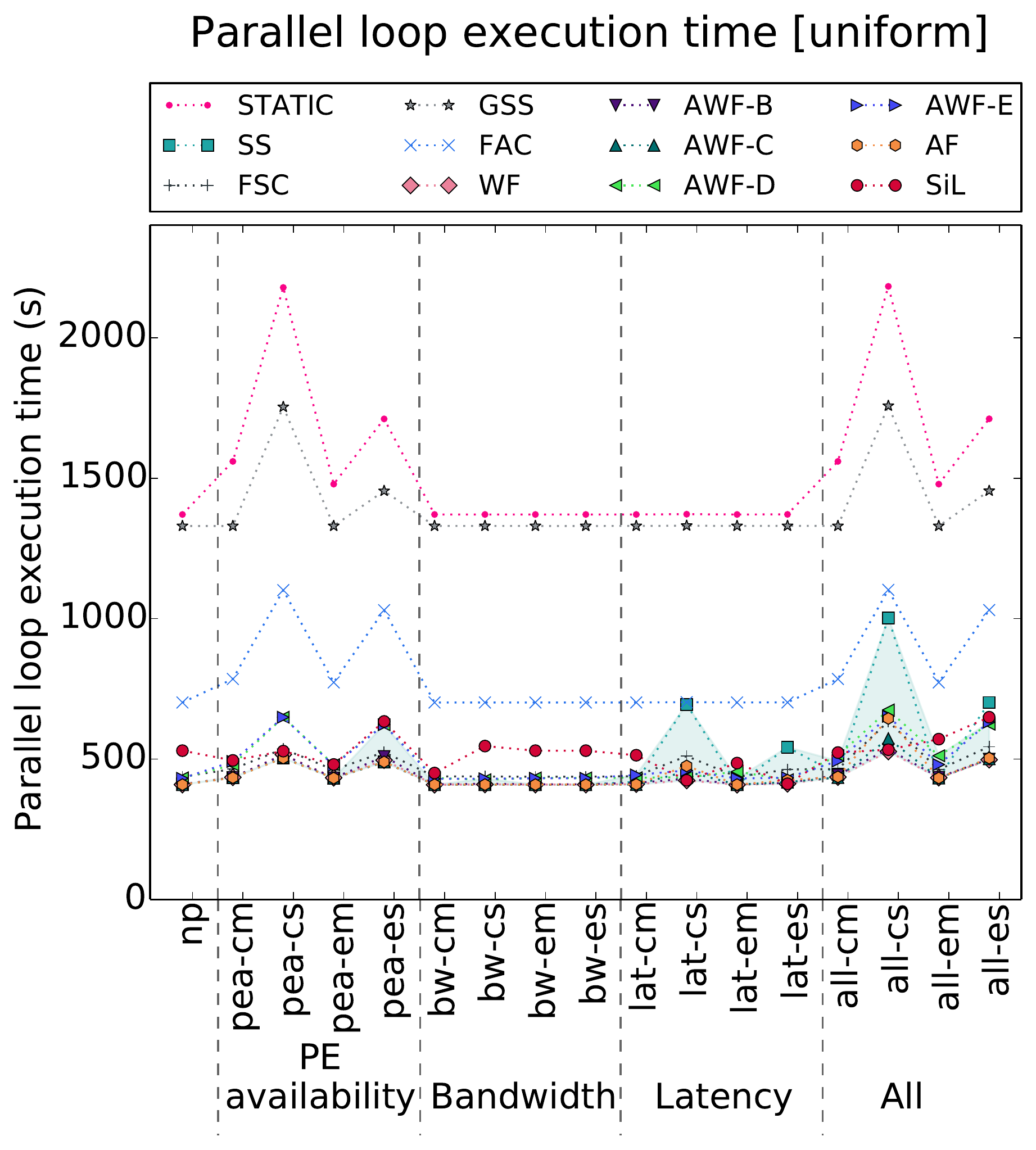}%
			\label{subfig:uniform}%
		} \hspace{0cm}
		\subfloat[Normal distribution on 696 cores ]{%
			\includegraphics[clip, trim=0cm 0.5cm 0cm 3.9cm,scale=0.36]{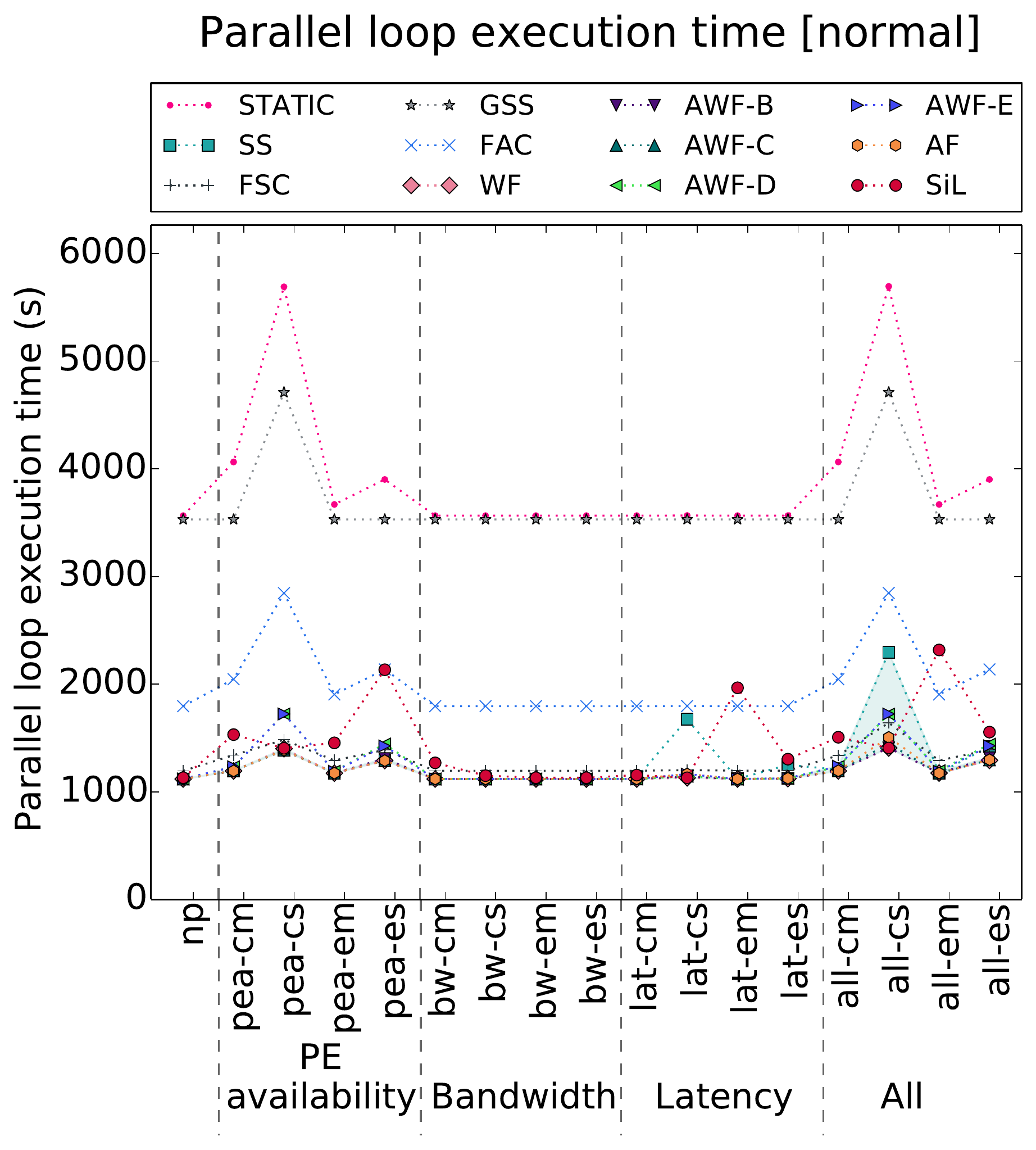}%%
			\label{subfig:normal}%
		} \vspace{-0.25cm}
		\\
		\subfloat[Exponential distribution on 696 cores]{%
			\includegraphics[clip, trim=0cm 0.5cm 0cm 3.9cm,scale=0.36]{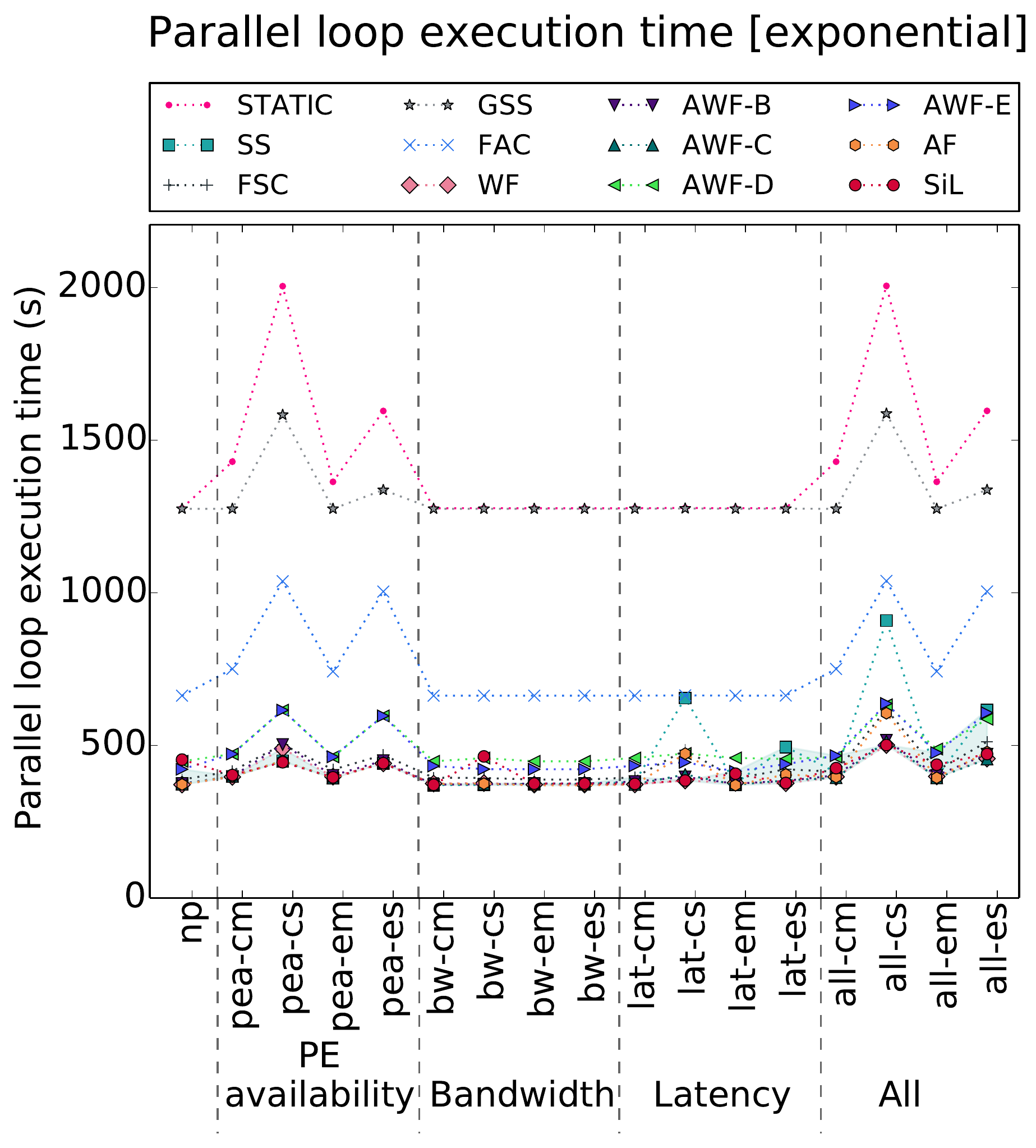}%
			\label{subfig:expo}%
		} \hspace{0cm}
		\subfloat[Gamma distribution on 696 cores]{%
			\includegraphics[clip, trim=0cm 0.5cm 0cm 3.9cm,scale=0.36]{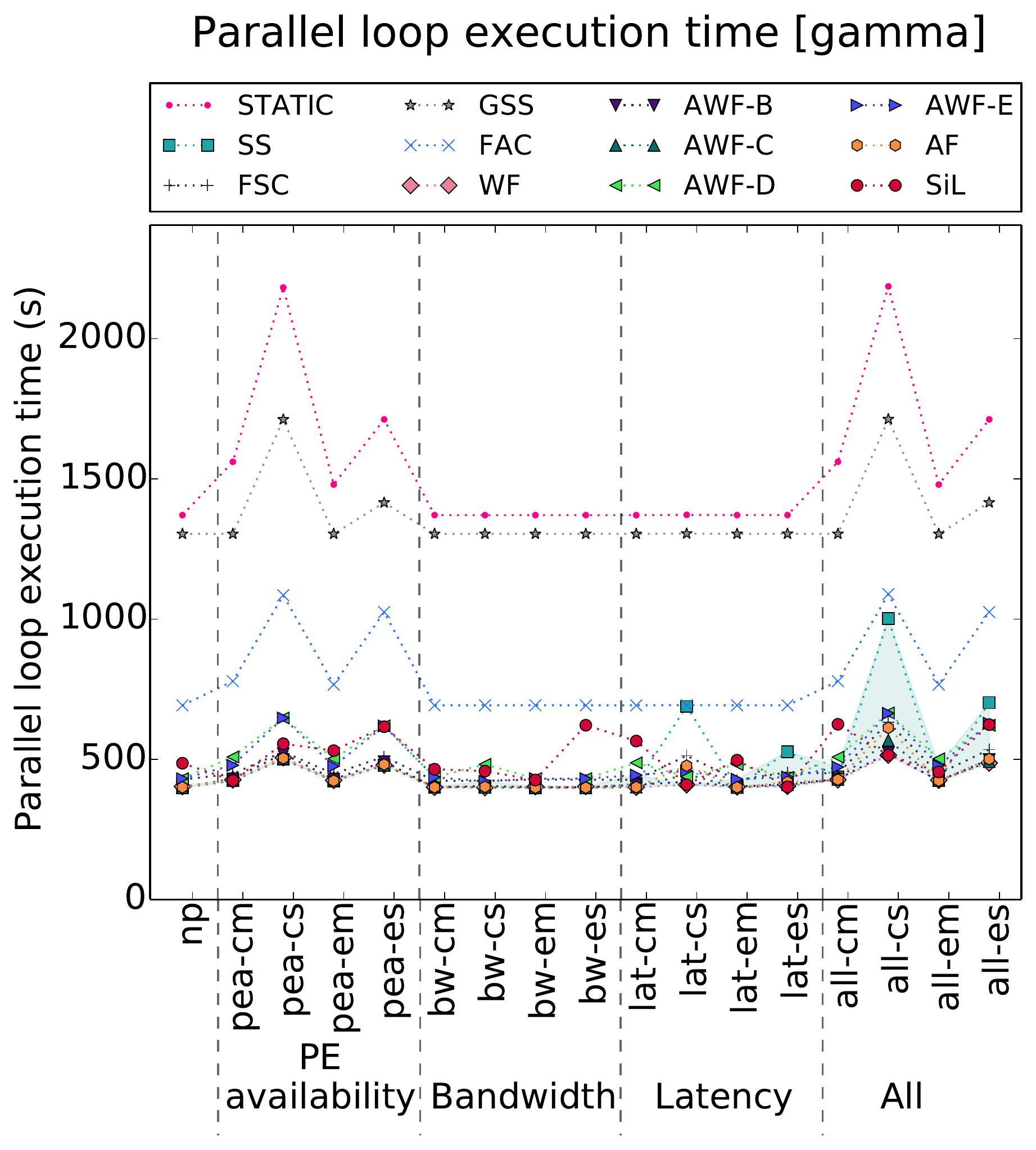}%%
			\label{subfig:gamma}%
		} \vspace{-0.35cm}
		\\
		\subfloat{%
			\includegraphics[clip, trim=0cm 16.9cm 0cm 2.1cm,scale=0.6]{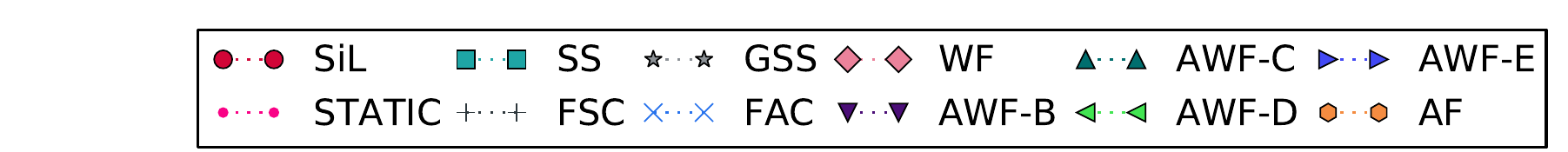}%
			\label{subfig:legend}%
		}
		\vspace{-0.25cm}
		%\end{adjustbox}
		\caption{Performance results of the six applications of interest without (np) and with (the rest) perturbations using SiL and eleven loop scheduling techniques on $696$ heterogeneous cores. The \aliA{mint color shaded regions denote} the upper and lower \aliA{bounds} of the performance with SiL if only one DLS technique were selected during execution \aliA{in the particular execution scenario}.} %\flo{Do not change the alignment in this figure. It should be OK as it is now.}}
		\label{fig:perf_dls}
	\end{minipage}
\end{figure}

\vspace{0.5cm}
\textbf{Performance of Scientific Applications under Perturbations.}
\label{subsec:perf}
%
%The six applications of interest are represented in the \simdag{} interface as described in~\ref{paragraph:app}.
%The miniHPC platform is represented as described in~\ref{paragraph:comp} and the perturbations are inserted using availability, bandwidth, latency files as described in~\ref{paragraph:per}.
%Results of other applications are available online\footnote{switch drive link to all raw data and figures}.\\
%
The performance of the six applications of interest is shown in \figurename{~\ref{fig:perf_dls}}.
One can see that STATIC, FSC, GSS, and FAC perform poorly on heterogeneous systems.
WF is well suited for scheduling on heterogeneous systems.
However, it can not adapt to accommodate the variability in the system due to perturbations, especially perturbations in the delivered computational speed.
SS is resilient to perturbations in the delivered computational speed of the PEs. 
However, it is significantly influenced by the network latency variations, as can be seen in \figurename{~\ref{subfig:PSIA}} ``lat-cs'' and ``lat-es''.
Perturbations in the network bandwidth show a very small influence on performance, as the PEs only communicate loop iterations indices to calculate the start index of the next chunk. 
Therefore, the communicated messages are small.

The adaptive techniques perform comparably, with a slight advantage for AWF-C as can be seen in \figurename{~\ref{subfig:expo}} ``all-cs'' and in \figurename{~\ref{subfig:PSIA}} ``pea-cs'' and ``all-es''.
%AWF-C outperforms all other techniques, in general.
However, in certain cases, other techniques outperform AWF-C. 
Specifically, WF outperforms AWF-C in \figurename{~\ref{subfig:PSIA}} ``lat-cs'' and ``all-cs''.
\emph{These results suggest that no single DLS outperforms all other techniques in all execution scenarios}. 
Therefore, the best strategy is to dynamically select a DLS based on the current application and system \aliA{states.}
 
\aliA{In this work,} SiL is called every 50~seconds to select the best performing DLS. %in the current execution scenario.
A closer analysis of the SiL-based results reveals that it resulted in the smallest execution time in most execution scenarios, especially for PSIA, as shown in~\figurename{~\ref{subfig:PSIA}}. 
The PSIA execution with SiL in the ``all-es'' scenario outperformed all other techniques, as the best DLS technique was changed during the execution according to the execution scenario.
In other cases, the application performance with SiL was slightly slower than the minimum execution time achieved by other DLS. 
This is due to the fact that loop scheduling is, by definition, non-preemptive and the execution of already scheduled loop iterations can not be preempted to be resumed with the newly selected DLS.
%\discuss{Even beyond this implementation, we may not arrive at preemption!}

\vspace{0.5cm}
\textbf{Discussion.}
%The performance of the six applications of interest with different DLS under various perturbations is discussed hereafter. 
\label{subsec:discussion}
\noindent%\discuss{The performance results suggest that no one single DLS technique achieves the best performance in all considered execution scenarios. \flo{This has been stated just above. Why repeat?}}
The advantage of  the SiL approach is to dynamically select the DLS that is predicted to achieve the best performance.
A combination of two or more DLS techniques throughout the application execution may result in \ali{a shorter} execution time than that achievable by any single DLS technique alone as can be seen in~\figurename{~\ref{subfig:PSIA} ``all-es''.
The SiL selected WF for the first 50 seconds in ``all-es'', as can be seen in~\figurename{~\ref{fig:sil_select}}. 
After 50 seconds, the network was no longer perturbed, and SiL selects the SS technique to balance the load and achieve a better performance than any single DLS technique.
%
%\FloatBarrier
\begin{figure}%{0.4\textwidth}
%	\vspace{-0.8cm}
	\centering
	\includegraphics[width=0.4\textwidth]{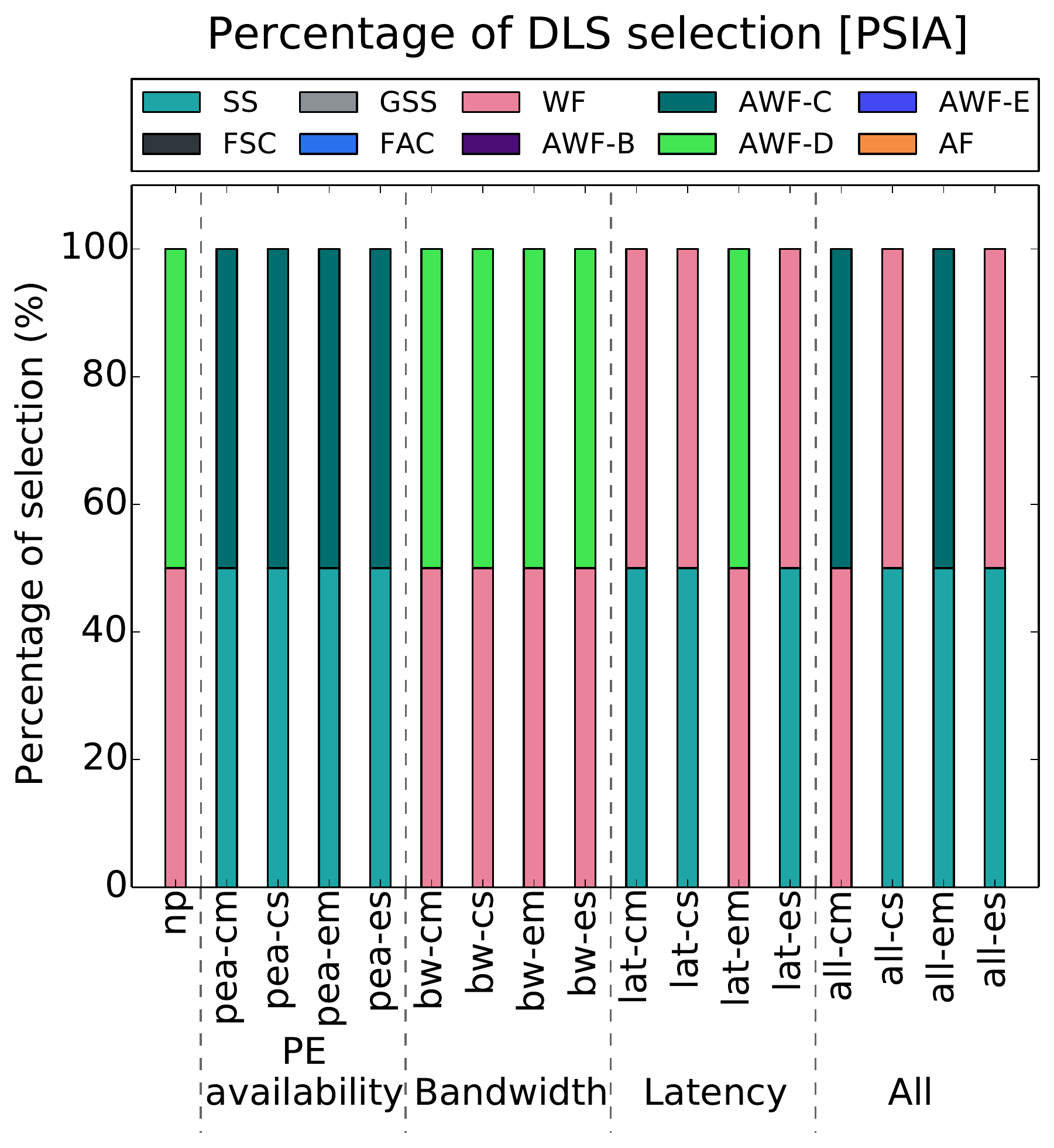}%
	\caption[caption of figure]{DLS selection results for the PSIA application. DLS techniques, such as FSC, GSS, and FAC are not selected due to their poor predicted performance with SiL.}
	\label{fig:sil_select}
%	\vspace{-0.5cm}
\end{figure}
\aliA{The simulative performance results of the PSIA on $224$~heterogeneous cores \aliA{(112 Broadwell cores and 112 KNL cores)} \cut{with adaptive and nonadaptive DLS techniques }have been verified by native experimentation under the \textit{no perturbation} execution scenario.}
The raw results and details of the DLS selection for all the applications can be found online\footnote{Included in this arxiv.org submission, please download all data}.
\aliA{The native experimentation of application performance in other execution scenarios is planned as immediate future work.}
In certain cases, such as ``all-em'' in the application with normally distributed tasks, the SiL-based execution did not yield the best performance, due to the \aliA{fact that DLS is} non-preemptive.
\aliA{The DLS techniques selected via SiL can be} used as guidelines \aliA{for a given} application, computing system, and perturbation scenario.
%The techniques that considers the assigning overhead such as AWF-D and AWF-E are selected more frequently in the execution scenarios that have perturbations in the network latency, and hence longer chunk assignment overhead.
%\discuss{*SiL can be used to detect anomalies in the performance of applications during execution. 
%A deviation from the predicted performance of the application corresponding to an undetected perturbation in the system. 
%In such a case, the SiL is required to re-evaluate the DLS selection decision.*}
%For example, perturbations in latency can be detected as the performance of SS is significantly influenced by it and call the SiL to select another DLS.}
The SiL approach can proactively \aliA{select the best suited DLS before any perturbations act on the system}, when perturbations can be predicted in advance.
The study and prediction of perturbations on HPC systems need further \aliA{examination}, as perturbations in HPC shared~resources are inevitable.
\cut{The SiL leverages the use of already developed simulators, instead of building new predicting or learning techniques.
The accuracy and the speed of the simulators will improve eventually with their development.} %THIS IS ALREADY SAID BEFORE - this is not the place to summarize, but to DISCUSS
The cost of the SiL simulation depends on the problem size and \aliA{the} system size.
\aliA{Specifically}, simulating the execution of 20,000~iterations on $9$~PEs with \simdag{} executing on an Intel~Broadwell~E5 processor, with \mbox{CentOS} 7.2~operating system, required $0.34$~seconds on average, whereas, it required $3.48$~seconds for simulating the execution of 200,000~iterations on the same number of PEs.
These costs can be amortized or entirely mitigated by calling the simulator asynchronously to the parallel loop execution.
\cut{Approaches to represent and experimentally verify the simulation of an application on an HPC system have been previously introduced~\cite{Ali_SC:17,Mohammed:2018a}.
The achieved percent error between native and simulative performance results of PSIA using \simdag{} was between $0.95\%$ and $2.99\%$~\cite{Mohammed:2018a}.} %AGAIN: These have been repeated. This is not the place to summarize the paper, but to DISCUSS the results!
 
%\input{5.tex}
%acknowledgement to Ahmed and PASC
% !TEX root =  2018_dls_perturbations.tex
\section{Conclusion and Future Work}
\label{sec:conc}
%\begin{itemize}
%	\item  What has been achieved by the current research
%	 \item Discuss and reiterate major advantages and drawbacks of the new solution 
%	 \item follow a + - + pattern
%	 \item  Future work
%\end{itemize}
A \aliA{new control-theoretic inspired approach}, namely simulator in the loop (SiL), was introduced to dynamically select a DLS that achieves the best performance, in an effort to answer the question of which DLS technique will achieve improved performance under unpredictable perturbations.
The performance of six applications is studied under perturbations and insights on the resilience of the DLS techniques to perturbations are provided. 
The performance results confirm the hypothesis that no single DLS technique can achieve the best performance in all the considered execution scenarios.
Using \aliA{the} SiL approach improved the performance of applications in most considered experiments.
\discuss{SiL leverages state-of-the-art simulators to select the DLS predicted to result in the best performance of an application under perturbations.}
%\discuss{The alternative of using machine learning for DLS selection may perform comparably to the SiL-based DLS selection. While machine learning models require training and learning either prior to execution or during previous time-steps in time-stepping applications~\cite{sukhija:2014:a,Boulmier:2017a}, 
The SiL can be asynchronously launched concurrently to the application execution.
The results show that in \aliA{the case of a system perturbed via multiple sources}, a combination of two or more DLS techniques may result in \aliA{improved} performance than that achievable by any single DLS alone, such as the performance of the PSIA in ``all-es'' execution scenario.
However, due to \aliA{applications being non-preemptively scheduled}, changing the used DLS during the execution \ali{may not result in the best performance}.
Further work is \aliA{planned} to realize and evaluate the performance of the SiL approach using native experimentation. 
Furthermore, experiments to investigate and enhance the performance of SiL, in terms of improving the DLS selection strategy and the period between SiL calls \aliA{also planned as future work.}

 {%to conclusion cite antony and nitin SiL cost to machine learing}

\bibliographystyle{splncs04}
\small \bibliography{citedatabase}
\end{document}